\documentclass[conference]{IEEEtran}
\usepackage{cite}
\usepackage{amsmath,amssymb,amsfonts}
\usepackage{algorithmic}
\usepackage{graphicx}
\usepackage{textcomp}
\usepackage{xcolor}
\usepackage{url}

\def\BibTeX{{\rm B\kern-.05em{\sc i\kern-.025em b}\kern-.08em
    T\kern-.1667em\lower.7ex\hbox{E}\kern-.125emX}}

\begin{document}

\title{GPU-Augmented OLAP Execution Engine: GPU Offloading}

\author{\IEEEauthorblockN{Il-Sun Chang}
\IEEEauthorblockA{\textit{Independent Researcher} \\
Daegu, Republic of Korea \\
kyou0072@gmail.com}
}

\maketitle

\begin{abstract}
Modern OLAP systems (e.g., Snowflake, BigQuery) have significantly resolved I/O bottlenecks through storage-compute separation and columnar storage structures. However, as data scale increases to the petabyte level, the CPU cost of sorting and joining (specifically Top-K selection and Join Probe) performed in the execution layer is emerging as a new bottleneck. This bottleneck is difficult to mitigate solely through hardware scaling and requires ``selective offloading'' that considers transfer, computation, and integration costs together. This paper proposes a hybrid architecture that augments rather than replaces the existing execution engine. The core principle involves offloading only specific primitives where the CPU experiences bottlenecks (Top-K selection and Key-based Join Matching/Probe) to the GPU within a vectorized execution pipeline. To suppress data movement costs, we adopt a Key-Only transfer (sending only Keys and Pointers) combined with Late Materialization. Furthermore, we introduce a ``Risky Gate'' (Risk-Aware Gating) that triggers offloading only in ``risk/gain intervals'' by considering transfer, kernel, and post-processing costs, input size, and candidate set complexity ($K, M$), rather than applying GPU acceleration to all queries. We observe consistent improvements in tail latency (P95/P99) under our gated policy compared to always-on offloading, based on PostgreSQL microbenchmarks and GPU proxy measurements. Consequently, this study quantifies not just that GPUs are fast, but ``when'' to turn them on, providing practical offloading policies to improve tail latency (P95/P99) and avoid mis-offloading in small-N intervals in problematic scenarios. This work extends the same risk-aware gating principle used in GPU-assisted measurement for query optimization (arXiv:2512.19750) to execution-layer OLAP primitives.
\end{abstract}


\section{Introduction}

\subsection{Background: The Shifted Bottleneck}
For decades, the primary cause of database performance degradation was disk I/O. However, storage offloading technologies such as Exadata SmartScan and in-memory computing have dramatically increased data access speeds. As a result, the performance bottleneck has shifted from storage to the execution layer. Today, the key constraint is not the speed of ``fetching'' data, but the speed at which the CPU ``processes (compares/sorts)'' the data.

\subsection{The Offloading Paradox}
We define this phenomenon as the ``Offloading Paradox.'' The storage layer filters and delivers data at line rate, but the execution engine (PGA/Compute Node) spends all its time sorting and joining the delivered data. This is a structural limitation that arises from the process of forcing high-dimensional data into a one-dimensional linear order.

\subsection{Augmented Approach}
This study does not discard the existing engine but ``augments'' it. Modern OLAP engines already use a Vectorized Execution model that processes columnar data in vector units \cite{boncz2005monetdb}. The GPU is not a heterogeneous device but a common accelerator capable of performing large-scale vector operations (selection/aggregation/matching), which can be integrated into the existing pipeline in a ``plug-in'' manner. However, GPU offloading is not always beneficial (due to transfer overheads, small $N$); therefore, offloading must be performed selectively only in risk/gain intervals. Our prior work introduced risk-aware gating for GPU-assisted measurement at the optimizer stage \cite{chang2025gace}; here we apply the same cost-accounted gating idea to execution-layer bottlenecks.

\section{Architecture: Hybrid Execution Engine}

\subsection{Design Philosophy}
The system consists of three components:
\begin{itemize}
    \item \textbf{Classical Host (CPU):} Handles data storage, scanning, I/O control, and final result generation.
    \item \textbf{GPU Coprocessor:} Handles Top-K selection, key-based matching/probe, and GPU-friendly aggregation.
    \item \textbf{Risky Gate:} Conditionally triggers the GPU path based on input size ($N$), transfer bytes, candidate set complexity ($K, M$), and estimated CPU costs.
\end{itemize}
The crucial point is that the GPU is not an ``always-on'' accelerator but is called restrictively by a policy device that includes cost accounting.

\subsection{Risky Gate: When to Turn on the GPU}
GPU offloading provides ``interval gain'' rather than ``always gain.'' Therefore, this paper defines the Risky Gate from the execution engine's perspective using the following signals:
\begin{itemize}
    \item \textbf{Input Size ($N$):} As $N$ becomes smaller, Host-to-Device (H2D) and Device-to-Host (D2H) transfers dominate, making the CPU path more advantageous.
    \item \textbf{Transfer Bytes ($B$):} The lower the key-only ratio (i.e., closer to full-row), the more disadvantageous the GPU path becomes.
    \item \textbf{Candidate Set Complexity ($K, M$):} As the number of conditions to evaluate ($K$) or the number of candidate sets ($M$) increases, CPU costs increase linearly, whereas GPUs gain relative advantages through parallelization.
    \item \textbf{Estimated CPU Cost ($\hat{C}_{cpu}$):} We estimate CPU processing costs using a simple cost model (e.g., $\alpha N \log N$ or $\alpha \cdot N \cdot K$).
\end{itemize}
The Gate compares $\hat{C}_{cpu}$ with the measured offloading cost $\hat{C}_{gpu} = (\text{Transfer} + \text{Kernel} + \text{Post-processing})$ and triggers the GPU path only when $\hat{C}_{cpu} - \hat{C}_{gpu}$ exceeds a threshold. This improves tail latency (P95/P99) by leaving fast queries alone and defending only against risky/bottleneck intervals.

\subsection{Key-Only Offloading Strategy}
To handle large-scale data, we adopt Key-Only offloading utilizing the characteristics of columnar storage \cite{abadi2008materialization}.
\begin{itemize}
    \item \textbf{Zero/Low-Cost Extraction:} In columnar DBs, sort/join keys are stored continuously, so key extraction overhead is low.
    \item \textbf{Transmission:} Only (Key, Pointer(RowID)) is sent to the GPU to minimize bandwidth and H2D costs.
    \item \textbf{Processing:} The GPU performs Top-K or Matching/Probe on the key vector and returns only sorted pointers (or a set of matched pointers).
    \item \textbf{Late Materialization:} The Host re-accesses the necessary columns only for the very few returned RowIDs to assemble the results.
\end{itemize}
This principle directly addresses the reality that ``transmission is the bottleneck'' and is a core design constraint determining the economics of offloading.

\section{Primitives for OLAP}

We focus on two primitives that are core bottlenecks in OLAP queries, rather than the ambiguous goal of accelerating sorting entirely.

\subsection{Ordering Primitive (Top-K Selection)}
Since the cost of Full Sort surges to $O(N \log N)$, we define LIMIT-based Top-K selection, which appears most frequently in execution engines, as the primary target for offloading. On the GPU, this consists of key-only vector transfer, partial selection (or radix-based selection), and returning pointers of the Top-K results.

\subsection{Matching Primitive (Key-based Join Probe)}
When random access and collision search in the Probe phase of a hash join become bottlenecks, the GPU can perform parallel probing using a key-only hash table (or bloom-filter/partitioned hash). The principle remains the same: ``Key-Only processing $\rightarrow$ Pointer return $\rightarrow$ Late Materialization on Host.''

\subsection{Collaboration with FPGA}
A configuration where FPGA/SmartNIC handles WHERE filters/simple aggregation (streaming) and the GPU handles global ordering/matching can be complementary rather than conflicting.

\section{Execution Trace Case Study}

\subsection{Experiment 1: GPU Offload Top-K Latency Analysis}

\subsubsection{Confirming the Bandwidth Wall}
GPU acceleration latency can be determined by host-device transfer costs rather than computation performance itself. To verify this, we measured GPU execution by separating it into Full-row transfer and Key-only transfer.
The transfer bottleneck is quantified by the median of 31 runs in the Stage-1 proxy measurement (Fig. 6). Full-row transfer latency increases sharply as $N$ increases, dominating the bottleneck in large-scale inputs. Conversely, Key-only transfer mitigates latency increase by reducing the transfer volume under the same conditions. This demonstrates the existence of an interval where ``transfer, not computation, is the bottleneck'' when applying GPUs.

\subsubsection{Small-N Interval Overhead and Necessity of Risky Gate}
In the Small-N interval, GPU offloading can result in a net loss. This is because GPU kernel launching and transfer overheads are dominant in absolute terms.
Fig. 1 shows that the ``Always GPU (No Guard)'' strategy worsens latency in Small-N, whereas applying the Risky Gate (e.g., if $N < 20k$ then CPU) eliminates Small-N overhead while maintaining GPU utilization in large $N$ intervals.

\begin{figure}[htbp]
\centering
\includegraphics[width=0.9\linewidth]{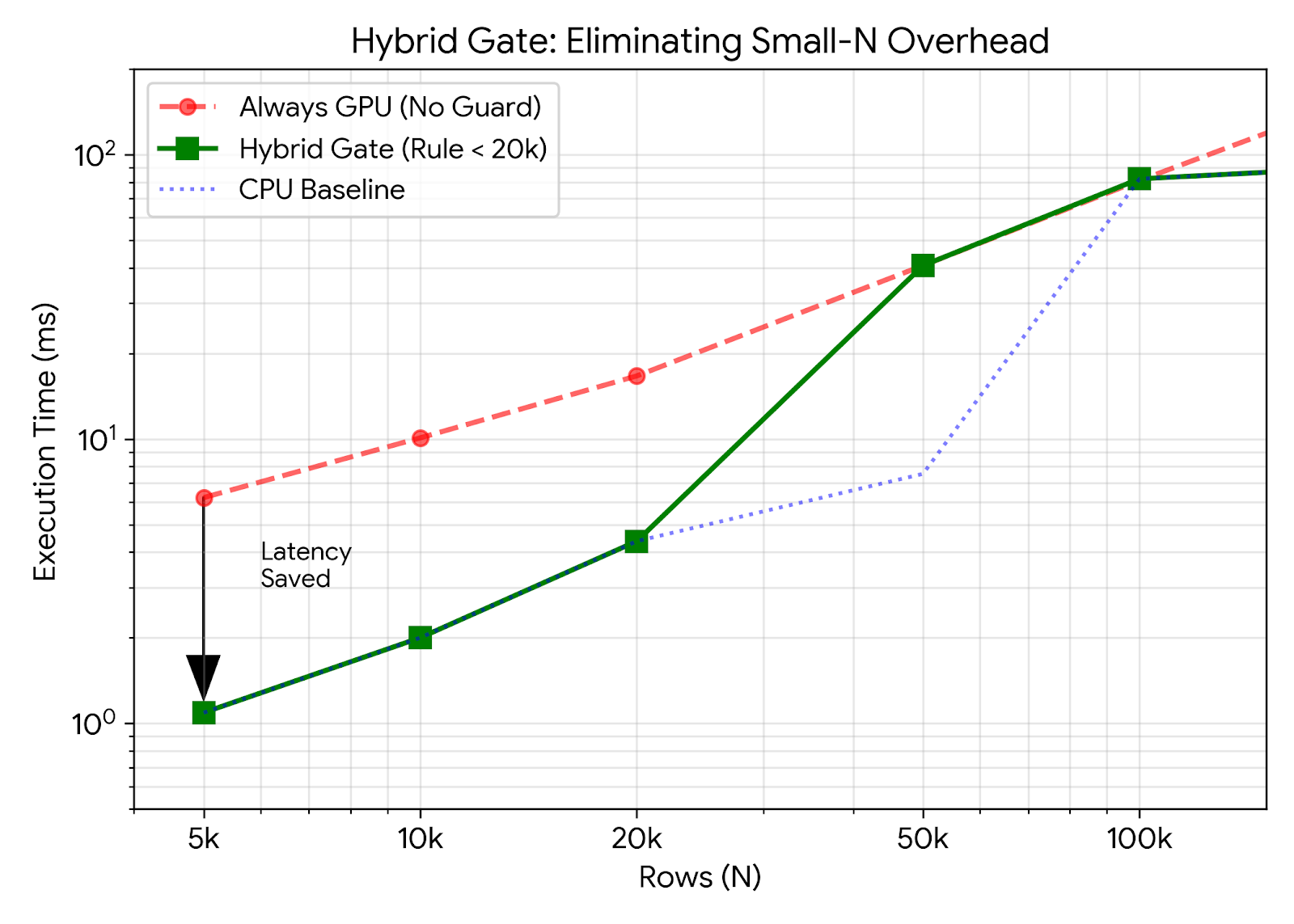} 
\caption{Risky Gate: Eliminating Small-N Overhead. The Risky Gate strategy avoids the latency penalty incurred by GPU offloading for small datasets.}
\label{fig1}
\end{figure}

\subsubsection{Offloading Sensitivity according to Gate Margin}
The Risky Gate should not be a fixed heuristic but a policy adjustable to system conditions. To this end, we measured the change in offloading ratio while varying the gate margin (ms). Fig. 2 shows that the threshold for switching to GPU shifts according to the margin value (0/5/10ms). This indicates that the offloading policy has a tuning knob capable of responding to workload/hardware changes.

\begin{figure}[htbp]
\centering
\includegraphics[width=0.9\linewidth]{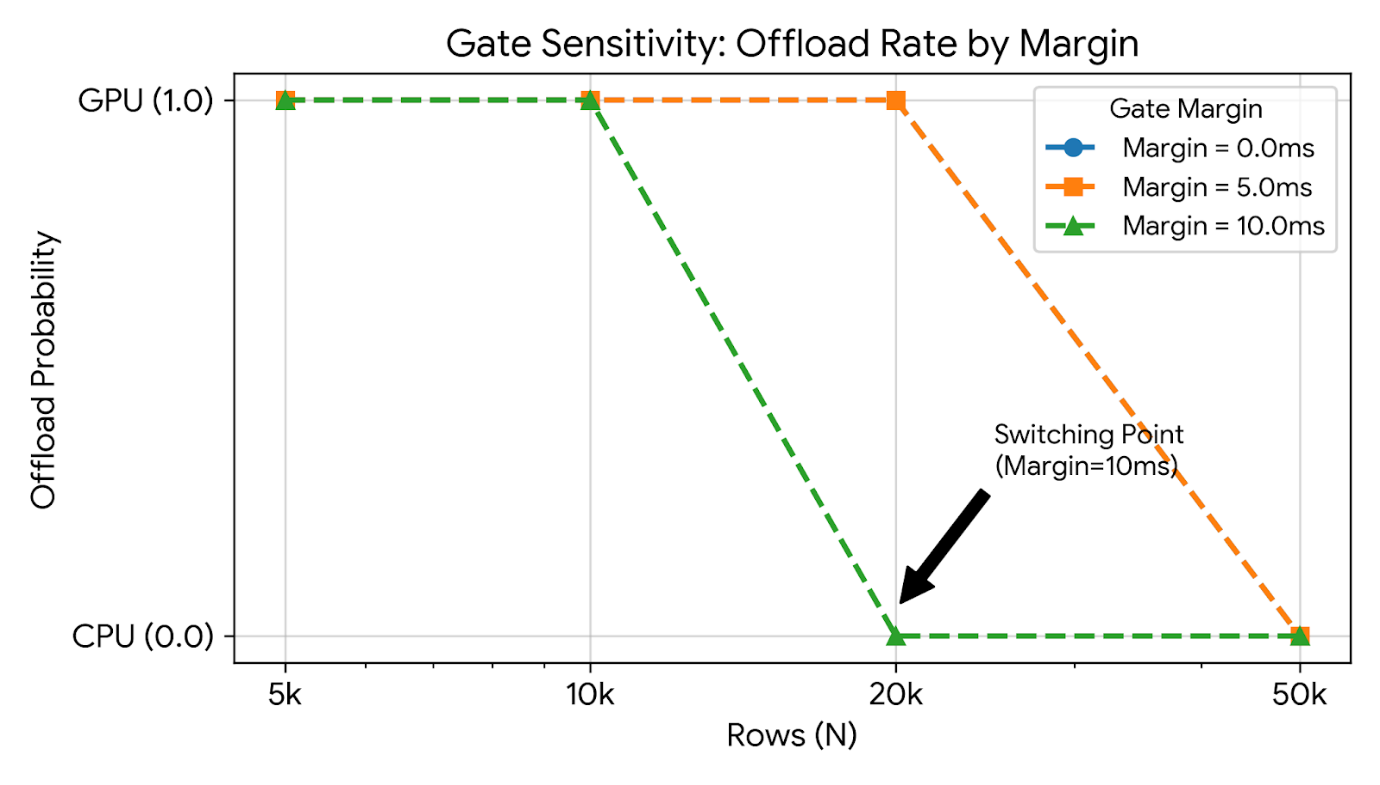} 
\caption{Gate Sensitivity: Offload Rate by Margin. The switching point shifts based on the defined margin, allowing for policy tuning.}
\label{fig2}
\end{figure}

\subsubsection{Summary of Experiment 1}
Experiment 1 confirmed that GPU offloading performance is significantly affected by the Bandwidth Wall and Small-N overhead rather than computation itself. Key-only transfer mitigates the transfer bottleneck, and the Risky Gate eliminates GPU misjudgments in the Small-N interval. Additionally, the gate margin operates as a parameter to control the offloading threshold.

\section{Performance Evaluation and Analysis}

In this study, to verify the utility of the proposed hybrid architecture, we established a performance prediction model based on benchmarking data from an actual commercial workstation.

\subsection{Experimental Setup}
We conducted PostgreSQL-based microbenchmarks to empirically justify the conditions of the offloading policy (when to offload). This experiment is not to claim GPU performance but to identify the growth rate of CPU execution costs, key-only transfer cost constants, and the break-even point ($N^*$) fitted to these constants.
\begin{itemize}
    \item \textbf{Host:} PostgreSQL 16, Single Workstation Environment
    \item \textbf{CPU:} AMD Ryzen 7 (8-Core)
    \item \textbf{RAM:} 64GB
    \item \textbf{GPU:} NVIDIA GeForce RTX 4060 Laptop GPU
    \item \textbf{Data:} Large-scale table containing value (double) and payload (text)
    \item \textbf{Stack:} PostgreSQL 16, Python, NumPy, PyTorch (GPU proxy)
    \item \textbf{Repetition:} REPEATS times for each $N$, reporting median and P95.
\end{itemize}
All experiments report median and P95 over repeated runs with a warmed cache, and the SQL/query templates and measurement scripts are available upon request.

\subsection{CPU Baselines: Full Sort vs Top-K Scaling}
In OLAP workloads, Full Sort has a theoretical lower bound of $O(N \log N)$ for both classical and GPU approaches; thus, we define Top-K selection as the main offloading target. To verify this, we measured the execution times of Full Sort and Top-K while increasing $N$.
\begin{itemize}
    \item Full Sort: \texttt{ORDER BY value}
    \item Top-K: \texttt{ORDER BY value DESC LIMIT 100}
\end{itemize}
Fig. 3 shows that Full Sort increases sharply with $N$, while Top-K shows a relatively moderate increase. This empirically supports the validity of the design choice to selectively offload only the Top-K primitive, rather than ``accelerating full sort.''

\begin{figure}[htbp]
\centering
\includegraphics[width=0.9\linewidth]{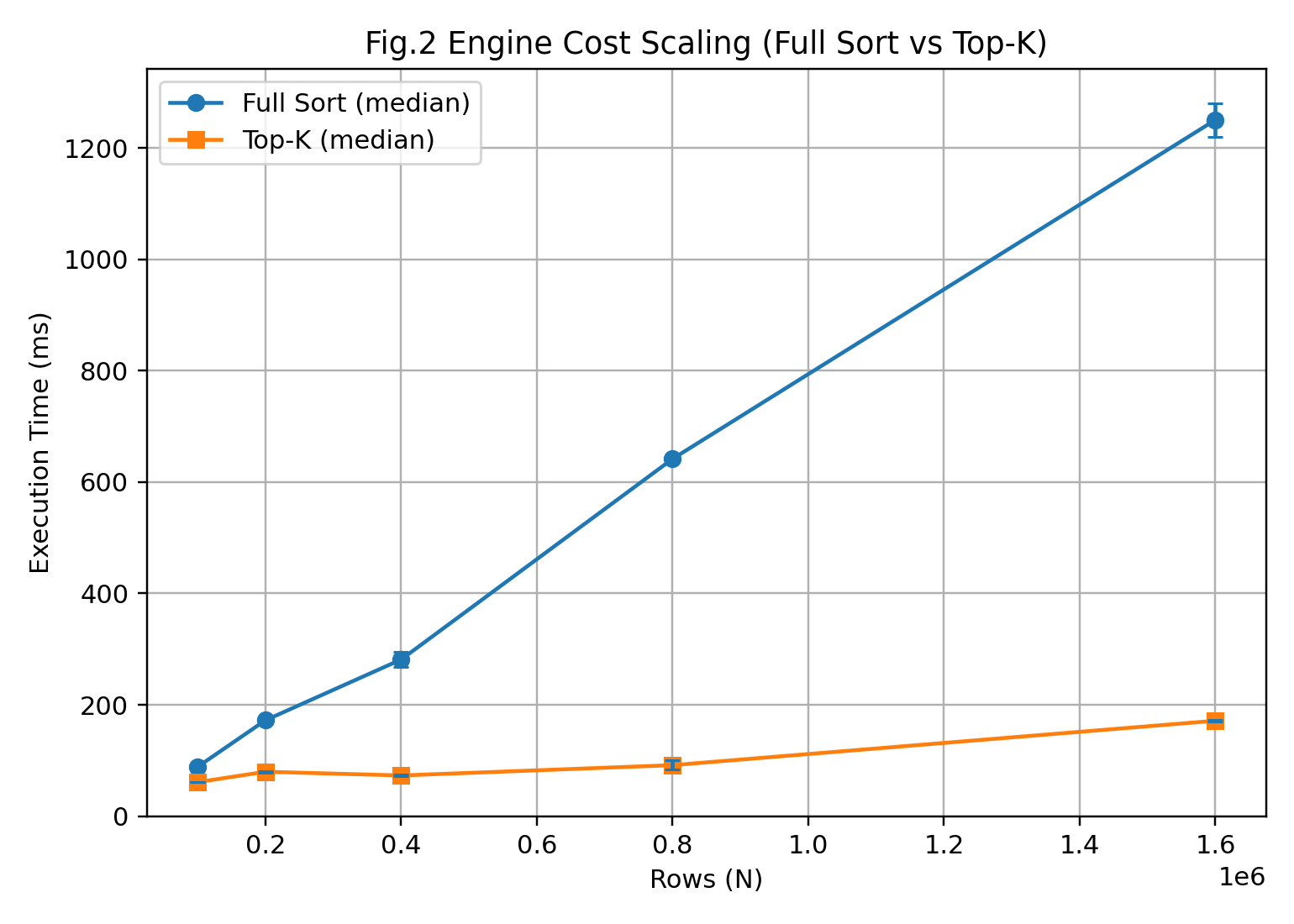} 
\caption{Engine Cost Scaling (Full Sort vs Top-K). Full Sort costs scale drastically ($N \log N$), whereas Top-K remains relatively low.}
\label{fig3}
\end{figure}

\subsection{Microbenchmarks: Key-only Extraction/Transfer Cost}
One of the core assumptions of this study is the necessity of Key-Only offloading. To verify this, we compared the fetch time of full-row transfer versus key-only (id, value) transfer from PostgreSQL using a Python client.
Fig. 4 shows that Key-only transfer exhibits consistently lower latency compared to Full-row transfer, and transfer bytes are also significantly reduced. This supports the practical reason for limiting the input delivered to the GPU to (Key, Pointer).

\begin{figure}[htbp]
\centering
\includegraphics[width=0.9\linewidth]{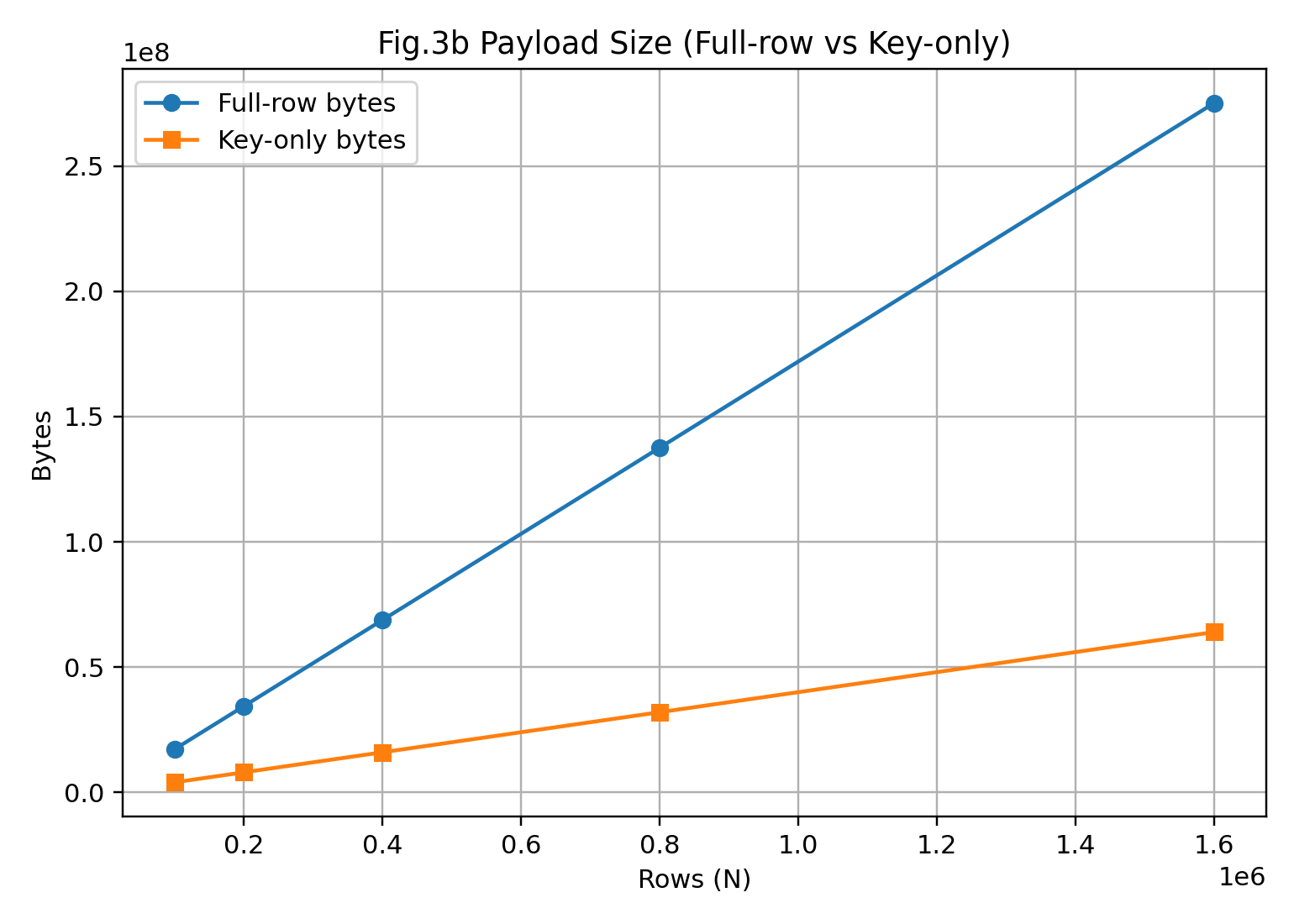} 
\caption{Payload Size (Full-row vs Key-only). Key-only transfer significantly reduces the data payload.}
\label{fig4}
\end{figure}

The empirical results from Sections 5.2 to 5.4 reaffirm that Full Sort surges as $O(N \log N)$ while Key-only transfer increases linearly as $O(N)$, quantitatively supporting the offloading conditions of this hybrid engine.

\subsection{Break-even Analysis: Measured Constants Fitted}
We fitted constants from the measurement results to reconstruct the Golden Cross as a result fitted to measured constants rather than assumptions.
\begin{itemize}
    \item CPU Full Sort Model: $T_{cpu}(N)=a\cdot N\log_{2}N+b$
    \item Key-only Transfer Model: $T_{tx}(N)=c\cdot N+d$
\end{itemize}
The break-even point $N^*$ is defined as the solution satisfying $T_{cpu}(N^{*})=T_{tx}(N^{*})+T_{gpu}(N^{*})$. Fig. 5 shows that the predicted $N^*$ matches the measured intersection point with a 2.7\% error. Note that $\hat{C}_{gpu}(N)$ includes the end-to-end costs (transfer + kernel + result return + Late Materialization) measured in the Stage-1 proxy.

\begin{figure}[htbp]
\centering
\includegraphics[width=0.9\linewidth]{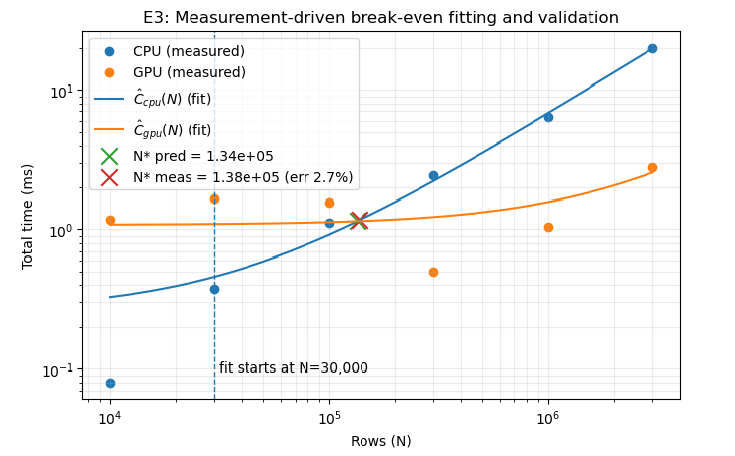} 
\caption{Measurement-driven break-even fitting and validation. The predicted break-even point closely matches the measured values.}
\label{fig5}
\end{figure}

\subsection{Stage-1 Proxy: GPU Top-K Microbenchmark}
This study reinforces the possibility of offloading from the execution engine's perspective by presenting the measured overhead of ``Transfer + Kernel + Result Return'' through a GPU-based Stage-1 proxy.
Fig. 6 and Fig. 7 show the magnitude of each cost and the E2E trend as data scale increases.
Full-row transfer costs increase sharply with $N$, whereas key-only transfer significantly reduces transfer costs by eliminating payload (16.2x reduction at $N=3M$). Key-only + late materialization significantly reduces total latency compared to full-row, showing a maximum speedup of 12.9x at $N=3M$.

\begin{figure}[htbp]
\centering
\includegraphics[width=0.9\linewidth]{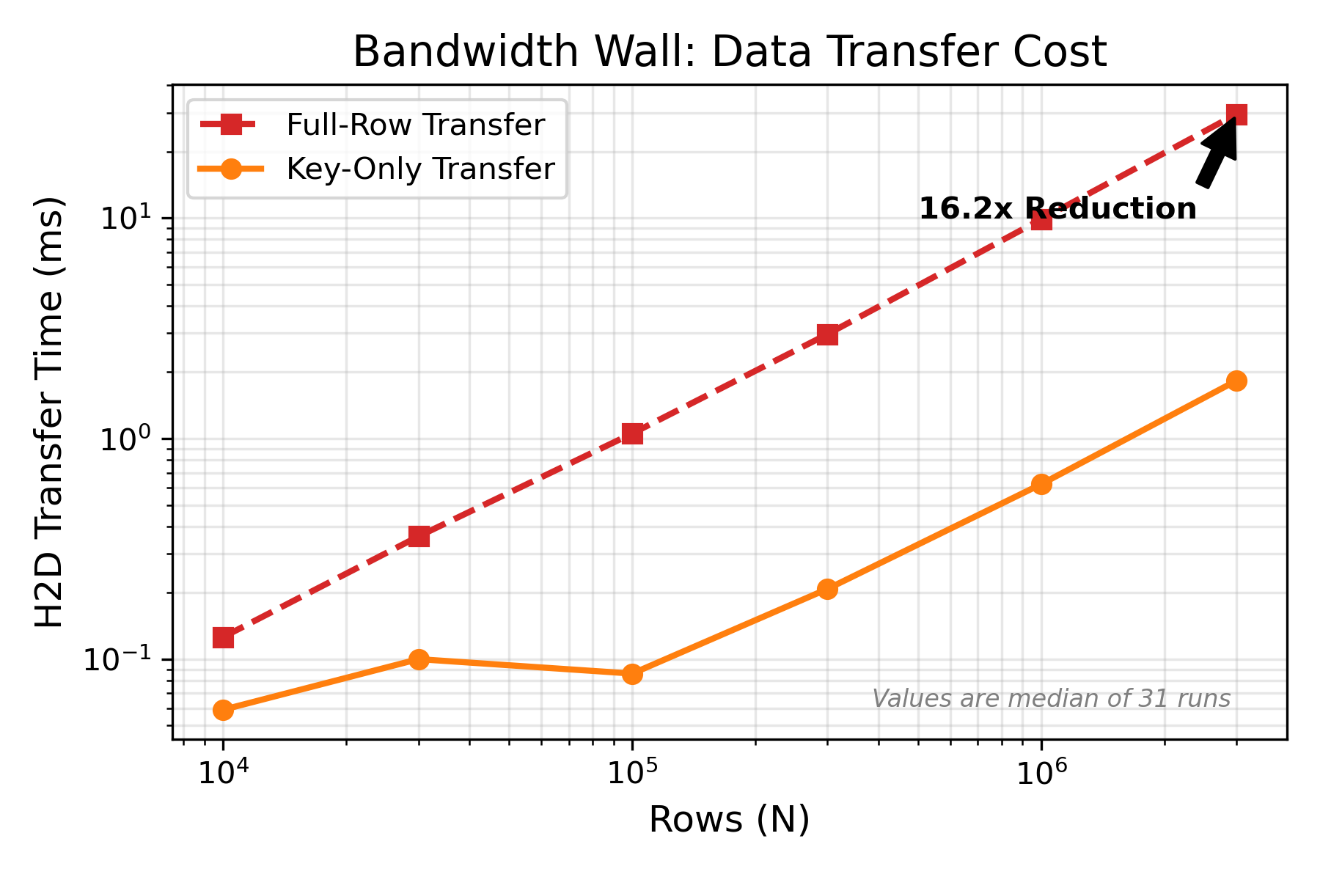} 
\caption{Bandwidth Wall: Data Transfer Cost. Key-Only transfer provides a 16.2x reduction in transfer time at $N=3M$.}
\label{fig6}
\end{figure}

\begin{figure}[htbp]
\centering
\includegraphics[width=0.9\linewidth]{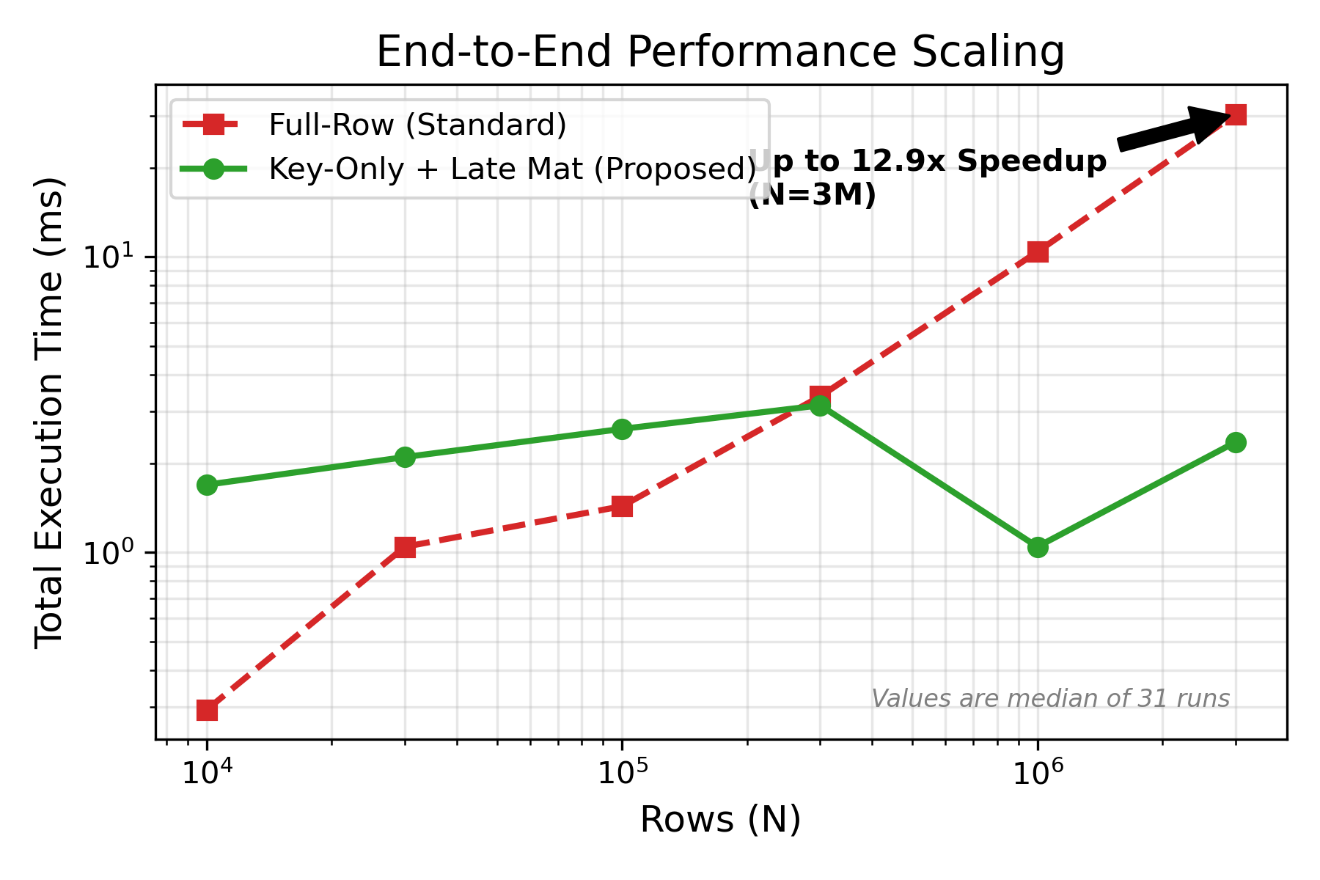} 
\caption{End-to-End Performance Scaling. The proposed Key-Only + Late Materialization method achieves up to 12.9x speedup.}
\label{fig7}
\end{figure}

\section{Conclusion}
This study proposed a hybrid execution engine that selectively offloads only high-impact primitives, such as Top-K selection and Key-based Matching/Probe, to GPUs to mitigate the processing barrier (sort/join) of the execution layer that emerged after I/O bottlenecks were resolved in modern OLAP systems. The core contributions are suppressing transfer costs through Key-Only + Late Materialization and quantifying ``when to turn on the GPU'' using a Risky Gate-based cost accounting policy, reflecting the reality that offloading is not always beneficial. In summary, this paper identifies bottleneck primitives, applies the Key-Only principle to avoid transfer-dominated intervals, and presents a practical path to improve execution robustness and tail latency through break-even point-based gating.

\section*{Acknowledgments}
The author used AI-assisted tools (e.g., ChatGPT) for English grammar correction and language polishing. All technical content, experiments, and conclusions were developed and verified by the author.


\begin{thebibliography}{00}

\bibitem{snowflake}
B. Dageville \textit{et al.}, ``The Snowflake Elastic Data Warehouse,'' in \textit{Proc. ACM SIGMOD Int. Conf. Manage. Data}, 2016, pp. 215--226.

\bibitem{boncz2005monetdb}
P. A. Boncz, M. Zukowski, and N. Nes, ``MonetDB/X100: Hyper-Pipelining Query Execution,'' in \textit{Proc. CIDR}, 2005, pp. 225--237.

\bibitem{abadi2008materialization}
D. J. Abadi, D. S. Myers, D. J. DeWitt, and S. R. Madden, ``Materialization Strategies in a Column-Oriented DBMS,'' in \textit{Proc. IEEE ICDE}, 2007, pp. 466--475.

\bibitem{bress2014gpu}
S. Bre\ss, N. Siegmund, L. Bellatreche, and G. Saake, ``The Design and Implementation of GPU-Accelerated Main-Memory Database Systems,'' in \textit{Proc. SSDBM}, 2014, Art. no. 12.

\bibitem{he2008relational}
B. He, K. Yang, R. Fang, M. Lu, N. Govindaraju, Q. Luo, and P. V. Sander, ``Relational joins on graphics processors,'' in \textit{Proc. ACM SIGMOD}, 2008, pp. 511--524.

\bibitem{exadata}
Oracle, ``Oracle Exadata Database Machine X8M,'' White Paper, 2019.

\bibitem{chang2025gace}
I. Chang, ``Risk-Aware GPU-Assisted Cardinality Estimation for Cost-Based Query Optimizers,'' arXiv preprint arXiv:2512.19750, 2025.

\end{thebibliography}
\end{document}